\documentclass[manuscript]{emulateapj}
\usepackage{natbib}
\usepackage{times}
\usepackage{amsmath}
\usepackage[usenames,dvipsnames]{color}
\usepackage{MnSymbol}
\usepackage{multirow}

\slugcomment{Accepted for publication in The Astrophysical Journal Letters}

\shorttitle{Unveiling a Rich System of Faint Dwarf Galaxies in the Next Generation Fornax Survey}
\shortauthors{Mu\~noz et al.}

\begin{document}

\title{Unveiling a Rich System of Faint Dwarf Galaxies in the Next Generation Fornax Survey}

\author{Roberto~P.~Mu\~noz$^{1}$, Paul~Eigenthaler$^{1}$, Thomas H.~Puzia$^{1}$, Matthew A.~Taylor$^{1,2}$, Yasna~Ordenes-Brice\~no$^{1}$ , Karla~Alamo-Mart\'inez$^{1}$, Karen X. Ribbeck$^{1}$, Sim\'on~\'Angel$^{1}$, Massimo Capaccioli$^{3}$, Patrick C{\^o}t{\'e}$^{4}$, Laura Ferrarese$^4$, Gaspar Galaz$^{1}$, Maren Hempel$^{1}$, Michael Hilker$^{5}$, Andr\'{e}s Jord\'{a}n$^{1}$, Ariane~Lan\c{c}on$^{6}$, Steffen Mieske$^{2}$, Maurizio Paolillo$^{7}$, Tom Richtler$^{8}$, Ruben S\'anchez-Janssen$^{4}$, \& Hongxin Zhang$^{1,9,10}$}
\affil{
$^{1}$Instituto de Astrof\'isica, Pontificia Universidad Cat\'olica de Chile, Av.~Vicu\~na Mackenna 4860, 7820436 Macul, Santiago, Chile\\
$^{2}$European Southern Observatory, 3107 Alonso de C\'ordova, Vitacura, Santiago\\
$^{3}$INAF-Astronomical Observatory of Capodimonte, Salita Moiariello 16, 80131, Naples, Italy\\
$^{4}$NRC Herzberg Astronomy and Astrophysics, 5071 West Saanich Road, Victoria, BC V9E 2E7, Canada\\
$^{5}$European Southern Observatory, Karl-Schwarzchild-Str. 2, D-85748 Garching, Germany\\
$^{6}$Observatoire astronomique de Strasbourg, Universit\'e de Strasbourg, CNRS, UMR 7550, 11 rue de l'Universite, F-67000 Strasbourg, France\\
$^{7}$Department of Physics, University of Naples Federico II, C.U. Monte Sant'Angelo, via Cinthia, 80126, Naples, Italy\\
$^{8}$Departamento de Astronom\'ia, Universidad de Concepci\'on, Casilla 160-C, Concepci\'on, Chile\\
$^{9}$Chinese Academy of Sciences South America Center for Astronomy, Camino EI Observatorio 1515, Las Condes, Santiago, Chile\\
$^{10}$CAS-CONICYT Fellow
}
\email{rmunoz@astro.puc.cl}

\begin{abstract}
We report the discovery of 158 previously undetected dwarf galaxies in the Fornax cluster central regions using a deep coadded $u, g$ and $i$-band image obtained with the DECam wide-field camera mounted on the 4-meter Blanco telescope at the Cerro Tololo Interamerican Observatory as part of the {\it Next Generation Fornax Survey} (NGFS).~The new dwarf galaxies have quasi-exponential light profiles, effective radii $0.1\!<\!r_e\!<\!2.8$ kpc and average effective surface brightness values $22.0\!<\!\mu_i\!<\!28.0$ mag arcsec$^{-2}$.~We confirm the existence of ultra-diffuse galaxies (UDGs) in the Fornax core regions that resemble counterparts recently discovered in the Virgo and Coma galaxy clusters.~We also find extremely low surface brightness NGFS dwarfs, which are several magnitudes fainter than the classical UDGs.~The faintest dwarf candidate in our NGFS sample has an absolute magnitude of $M_i\!=\!-8.0$\,mag.~The nucleation fraction of the NGFS dwarf galaxy sample appears to decrease as a function of their total luminosity, reaching from a nucleation fraction of $>\!75\%$ at luminosities brighter than $M_i\!\simeq\!-15.0$ mag to $0\%$ at luminosities fainter than $M_i\!\simeq\!-10.0$ mag.~The two-point correlation function analysis of the NGFS dwarf sample shows an excess on length scales below $\sim\!100$\,kpc, pointing to the clustering of dwarf galaxies in the Fornax cluster core.
\end{abstract}


\section{Introduction}
The Fornax cluster is the most nearby southern galaxy overdensity and an excellent laboratory for studying the formation and evolution of galaxies and other dense stellar systems, such as globular clusters (GCs) and dwarf galaxies.~In comparison to its northern counterpart, the Virgo galaxy cluster, the Fornax cluster has twice the central galaxy density, half the velocity dispersion, and a distinctly lower mass \citep[$7\!\pm\!2\!\times\!10^{13}\,M_\odot$, e.g.][]{sch10}.~Furthermore, the core of Fornax is dynamically more evolved \citep{chu08} and its early-type galaxy fraction is significantly larger than that of Virgo.~Overall, these properties make the Fornax cluster an excellent target for studies of galaxy evolution processes in dense cluster environments.

Probing deeper into the galaxy luminosity function allows us to put more stringent constraints on galaxy formation models~\citep[e.g.][]{bov09, bov11a, bov11b, bro11, pfe14, vog14, mis15, bur15}.~In addition to the rich populations of low-surface brightness dwarf galaxies in the Local Volume \citep{mcc12}, there have been recent and surprising discoveries of ultra-diffuse galaxies (UDGs) in the Coma and Virgo galaxy clusters \citep{dokkum15a, koda15, mih15}.~The extremely low stellar masses of UDGs ($\sim\!6\!\times\!10^7\,M_\odot$) combined with their large radii ($1.5\!-\!4.6\,{\rm kpc}$) translate into very low surface brightness values in the range $\mu_{0,V}\!\approx\!26.0\!-\!28.5$ mag arcsec$^{-2}$, making them extremely difficult to detect.~The existence of this mysterious new dwarf galaxy class in dense galaxy cluster environments prompts the obvious question of whether there might be similar populations in other galaxy clusters.~In this letter we use multi-passband, wide-field observations conducted by us with the {\it Dark Energy Camera} \citep[DECam,][]{fla15} at the 4-meter Blanco telescope at Cerro Tololo Interamerican Observatory as part of {\it The Next Generation Fornax Survey} (NGFS) to search for faint dwarf galaxies in the Fornax galaxy cluster central regions ($R_{\rm NGC1399}\la350$\,kpc) and measure their structural parameters and spatial distribution.

Throughout this work, we adopt a Fornax distance modulus of $31.51$ mag ($20.0$\,Mpc) based on the surface-brightness fluctuations method from \cite{bla09}.

\begin{figure*}[!t]
\centering
\includegraphics[width=0.85\textwidth]{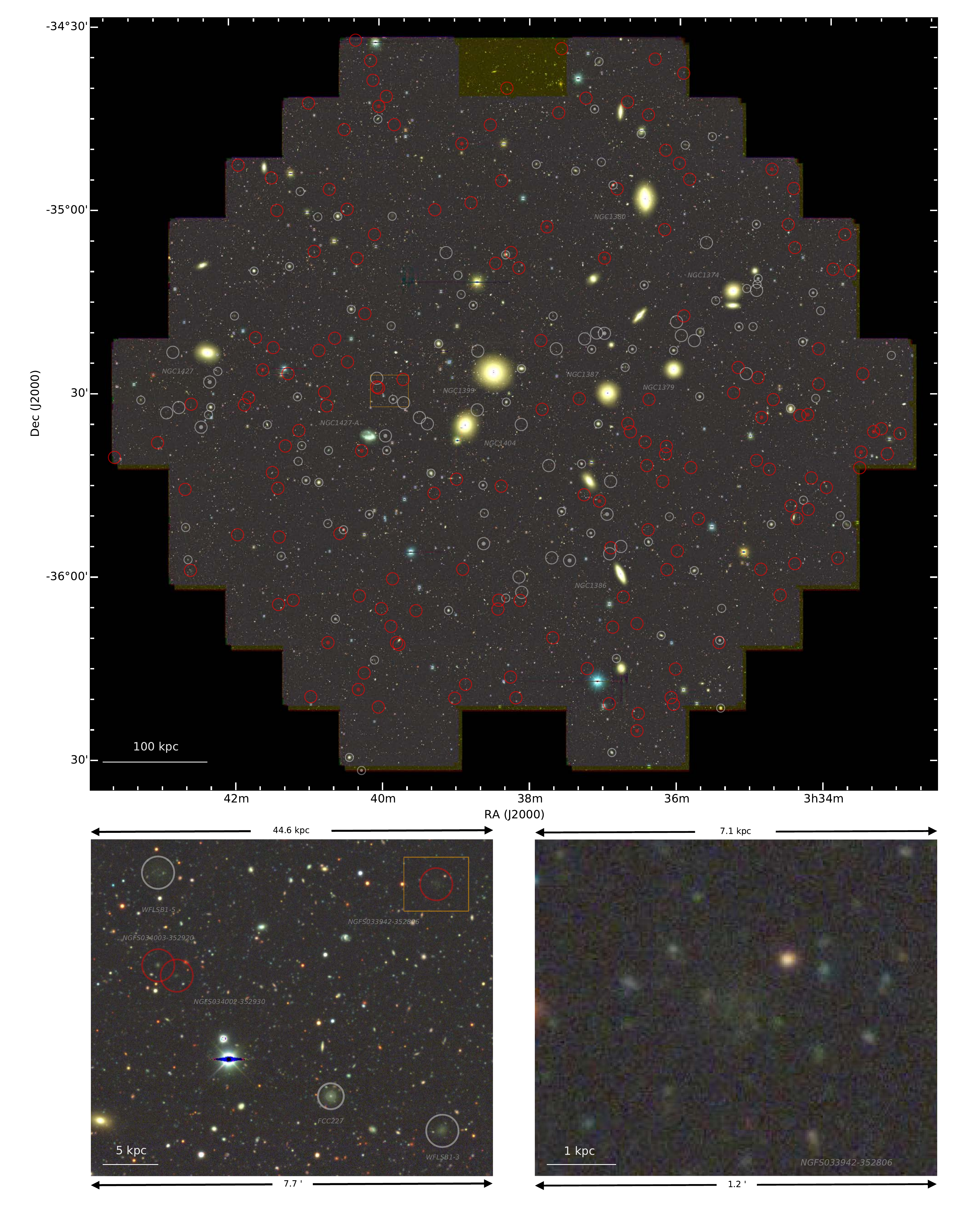}
\caption{({\it Top panel}): Illustration of the inner 3\;deg$^2$ of the NGFS survey footprint, centered on the brightest Fornax galaxy NGC\,1399, located close to the center of the field.~The image is a $ugi$ color composite stretched to accentuate the newly detected low-surface brightness dwarf galaxies, which are marked by red circles.~Small grey circles are dwarf galaxies identified by \cite{ferguson89} and listed in the FCC catalog. Large grey circles are additional galaxies identified by \cite{mie07}.~Nucleated dwarf galaxies are indicated by a central dot of the same color as the corresponding circle.~Note the striking grouping of some of the dwarf galaxies in the field.~The orange box shows the zoom-in area that is illustrated in the bottom left panel.~({\it Bottom left panel}): Detailed view of an exemplary region of the field where a variety of extended galaxies are located, among which is FCC\,227 and several compact and extended dwarf galaxy candidates in our NGFS data.~({\it Bottom right panel}): Zoom on one ultra-low surface brightness dwarf galaxy candidate, indicated by the orange box in the bottom left panel.}
\label{fig1}
\end{figure*}

\begin{deluxetable*}{cccccccccc}
\tablecaption{Dwarf galaxies in the core of the Fornax Cluster\label{tab:dwarftable}}
\tablehead{                                                                                                                                                                                                                                                                                                                               
\colhead{\multirow{2}{*}{ID}}       &   \colhead{\multirow{2}{*}{$\alpha_{2000}$}} &   \colhead{\multirow{2}{*}{$\delta_{2000}$}} &   \colhead{$m_i$}                 &  \colhead{$M_{i}\,$\tablenotemark{a}}     &  \colhead{\multirow{2}{*}{$n$\tablenotemark{b}}}   &  \colhead{${r_{{\rm{eff}}}}$}  &  \colhead{${r_{{\rm{eff}}}}$\tablenotemark{a}}   &   \colhead{\multirow{2}{*}{Type\tablenotemark{c}}}  &   \colhead{\multirow{2}{*}{Reference}} \\      
                                    &                                              &   \                                          &   \colhead{[mag]}               &  \colhead{[mag]}                          &                                   &  \colhead{[arcsec]}            &  \colhead{[kpc]}                &                                       &                                                 }           
\startdata                                                                                                                                                                                                                                                                                                                                                                                                          
NGFS033301-353635 & 03:33:00.83 & $-$35:36:34.59 & 17.70 & $-$13.81 & 0.89 & 5.397 & 0.523 & $\medcircle$ &  \\
NGFS033309-352349 & 03:33:08.63 & $-$35:23:49.02 & 19.18 & $-$12.33 & 0.81 & 7.088 & 0.687 & $\medcircle$ & FCC114 \\
NGFS033311-353956 & 03:33:10.93 & $-$35:39:56.17 & 20.56 & $-$10.95 & 0.63 & 8.950 & 0.868 & $\medcircle$ &  \\
NGFS033316-353551 & 03:33:15.98 & $-$35:35:50.90 & 21.00 & $-$10.51 & 0.69 & 2.554 & 0.248 & $\medcircle$ &  \\
NGFS033322-353620 & 03:33:22.18 & $-$35:36:20.20 & 18.70 & $-$12.81 & 0.88 & 4.439 & 0.430 & $\odot$ &  \\
NGFS033331-352654 & 03:33:31.41 & $-$35:26:54.00 & 22.74 & $-$8.77 & 0.47 & 2.427 & 0.235 & $\medcircle$ &  \\
NGFS033332-353942 & 03:33:32.11 & $-$35:39:42.21 & 19.86 & $-$11.65 & 1.00 & 8.258 & 0.801 & $\odot$ &    \\\vspace{-0.25cm}
\enddata
\tablenotetext{a}{Assuming a distance modulus of $(m\!-\!M)_0\!=\!31.51$ mag \citep{bla09}.}
\tablenotetext{b}{S\'{e}rsic index \citep{ser68, cao93}.}
\tablenotetext{c}{Morphological galaxy type classification: $\odot$=nucleated, $\medcircle$=non-nucleated dwarf galaxy.}
\tablecomments{Table~\ref{tab:dwarftable} is published in its entirety in the electronic edition of the {\it Astrophysical Journal}. A portion is shown here for guidance regarding its form and content.}
\end{deluxetable*}

\section{Observations and Image Processing}
\label{sec:observations}
The {\it Next Generation Fornax Survey} (NGFS) is an ongoing multi-wavelength survey of the central 30\;deg$^2$ of the Fornax galaxy cluster \citep{ferguson89}. We are using Blanco/DECam for doing near-ultraviolet (NUV) and optical photometry as well as VISTA/VIRCam \citep{sut15} for doing near-infrared (NIR) photometry of the same field.~The NGFS survey is aimed at detecting point-sources at S/N=5 over the PSF area with $u\!=\!26.5, g\!=\!26.1, i\!=\!25.3$, and $K_s\!=\!23.3$ AB mag. Below we offer a summary of the NGFS features most salient for this letter, and defer a detailed description of the characteristics and data reduction to a subsequent paper.

The current NGFS mosaic consists of nine contiguous tiles, and the present work is based on the optical imaging of the central tile. Our dithering pattern is based on the Elixir-LSB technique developed by Jean-Charles Cuillandre for the {\it Next Generation Virgo Survey} \citep[NGVS; see][]{ferrarese12}, which is designed to detect low-surface brightness structures.~The raw images were processed by the DECam Community Pipeline (v2.5.0), including bias calibration, crosstalk correction, linearity correction, flat fielding and gain calibration.~We apply our own sky background subtraction and stacking recipes via an iterative procedure of object masking and sky modelling using the {\sc Astromatic} software\footnote{http://www.astromatic.net/software} ({\sc Sextractor} v2.19.5, {\sc Scamp} v2.0.1 and {\sc Swarp} v2.38.0).~Our astrometric and photometric calibrations are based on 2MASS Point Source Catalog \citep[][]{skrutskie06} reference stars and Sloan Digital Sky Survey stripe 82 standard star frames taken under photometric conditions, respectively.

In this letter we present the analysis of the central 3\;deg$^2$ (Figure~\ref{fig1}), covering the Fornax cluster core out to $\sim\!350$\,kpc radius from NGC\,1399, corresponding to 25\% of the Fornax cluster virial radius \citep[1.4\,Mpc;][]{dri01b}.~We created a RGB image from the processed images, which was visually inspected by several of us (RPM, PE, THP, YO, KAM, KXR) to independently identify low-surface brightness dwarf galaxy candidates.~Table~\ref{tab:dwarftable} includes 284 dwarf galaxy candidates that were identified by every person on the search team, 264 of which we derived structural parameters for by analyzing their surface brightness profiles (see \S\ref{sec:structpars}).

We compared our sample with objects flagged as ``likely members" in the FCC catalogue \citep{ferguson89}, finding 92 matches with a maximum separation of $12\arcsec$.~Out of these galaxies, there are 47 matched non-nucleated and 45 nucleated dwarfs.~We also find 27 non-nucleated and 7 nucleated dwarfs not present in the FCC catalogue, but flagged as likely members in the \cite{mie07} catalogue.~We fail to recover 6 objects from the \citeauthor{mie07}~catalogue, which we visually confirm from Figure~\ref{fig1} to be background galaxies.

\section{Structural parameters of the dwarf candidates}
\label{sec:structpars}

We estimate the structural parameters of each dwarf candidate in Table~\ref{tab:dwarftable} using the software package {\sc galfit} \citep[v3.0.5;][]{pen10}.~We approximate one-component fits to the 2-dimensional galaxy surface brightness distribution assuming a S\'{e}rsic profile \citep{ser68, cao93}.~We split the fitting procedure into several steps: We create individual postage stamp images for each dwarf candidate and construct their segmentation maps using {\sc SExtractor} \citep{ber96}.~These segmentation maps are then used to create bad pixel masks for each dwarf, masking any non-dwarf sources above a 3$\sigma$ threshold, that produce dwarf-only images on which we perform the model fits.~We then characterize the dwarfs as either nucleated or non-nucleated.

\begin{figure*}[!t]
\centering
\includegraphics[width=\textwidth]{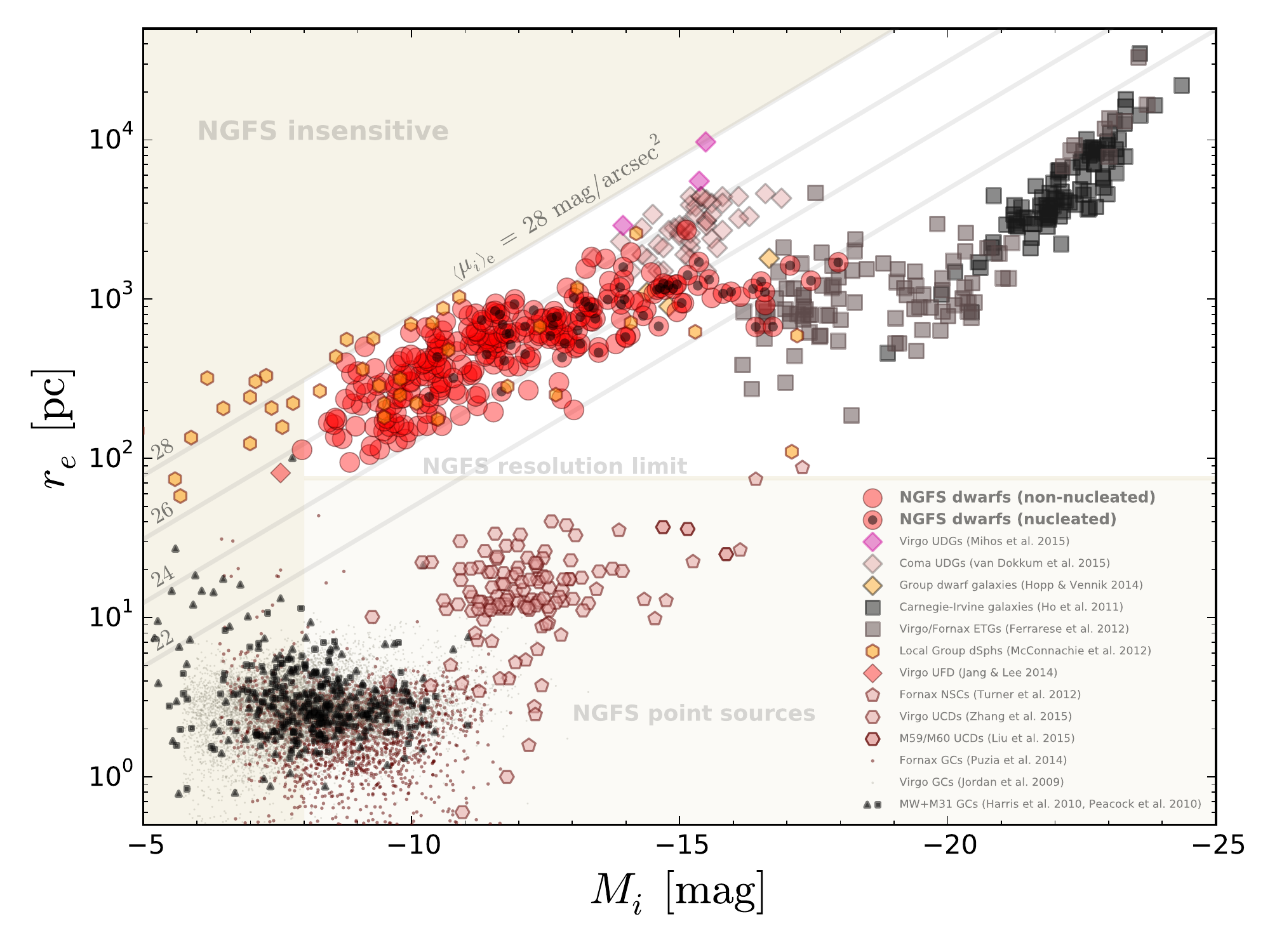}
\caption{Illustration of the size-luminosity relation, i.e.~effective radii vs.~absolute $i$-band magnitudes, for NGFS dwarf galaxies in the central regions of the Fornax galaxy cluster (see Fig.~\ref{fig1}) and other stellar systems in the nearby universe (see legend at the bottom right).~Red circles with black dots nucleated while open red circles are non-nucleated NGFS dwarfs.~Note that while nucleated dwarfs predominantly occupy a brighter luminosity range of the dwarf sequence, their size-luminosity relation in a given magnitude range is similar to that of the non-nucleated dwarfs, which are predominantly found at fainter luminosities.~Lines of constant average effective surface brightness are shown for $\left\langle \mu_i \right\rangle_{\rm{e}}\!=\!m_{\rm{tot}}\!+\!2.5\,\log(2\pi r_{\rm{e}}^2) =28, 26, 24, 22$ mag arcsec$^{-2}$. The shaded region to the top left marks the approximate surface brightness limit of our NGFS data and the point-source detection limit to the left. Objects smaller than the resolution limit appear as point in our NGFS data as indicated by the lightly-shaded region.~Note that the samples plotted in this figure have their own particular selection functions which naturally lead to the unphysical gaps, for instance, between NGFS dwarfs and brighter early-type galaxies as well as GCs and UCDs and NSCs.}
\label{dwarfsequence}
\end{figure*}

\begin{figure*}[!t]
\centering
\includegraphics[width=8.5cm]{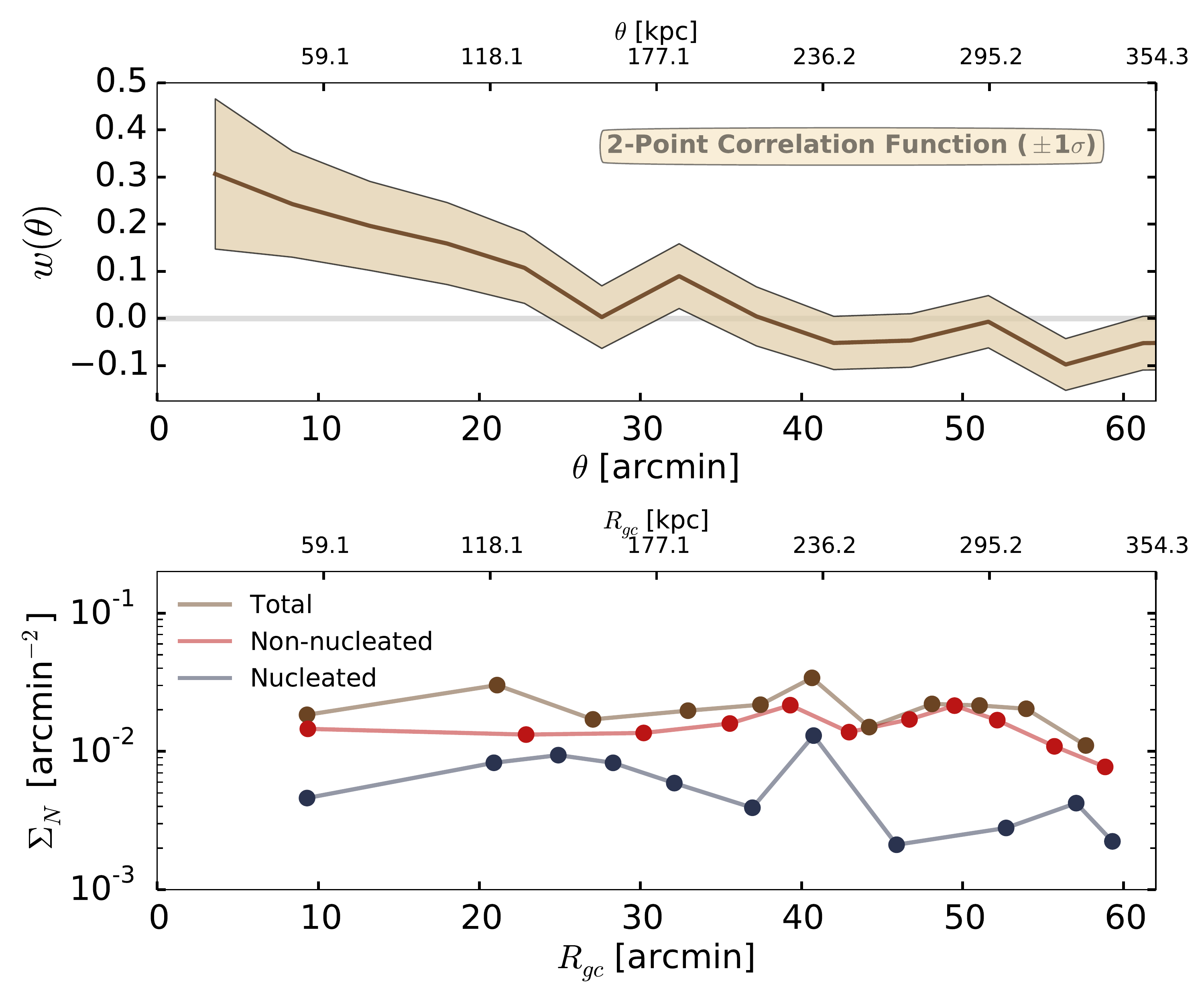}
\includegraphics[width=8.5cm]{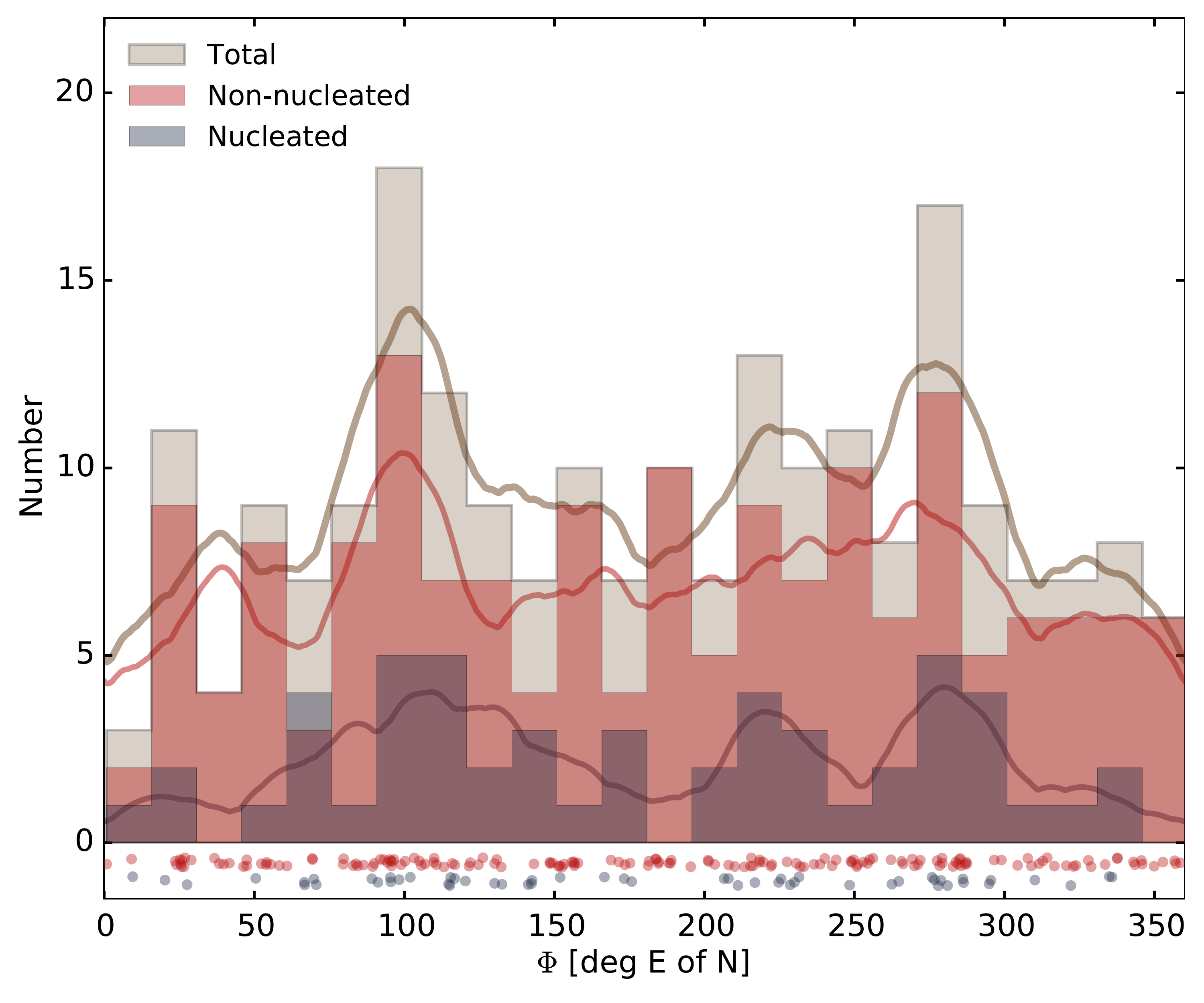}
\caption{The spatial distribution of the Fornax dwarf galaxy population. ({\it Top left panel}): The two-point correlation function is shown with 1$-\sigma$ bounds which indicates dwarf galaxy clustering below $\sim\!100$\,kpc scales. ({\it Bottom left panel}): The surface number density profile of dwarf galaxies is shown as a function of distance from NGC\,1399 for the non-nucleated (blue symbols), the nucleated (red), and the total (brown) samples in units of galaxies per arcmin$^2$.~Physical and angular units are shown on the top- and bottom axes, respectively. ({\it Right panel}): The azimuthal distribution of the three samples, in units of degrees East of North, is shown with the same color scheme as the bottom left panel. All samples show a general East-West bimodality, with discrete peaks in few preferred directions.}
\label{surfdense}
\end{figure*}

The non-nucleated dwarfs are fit by a single S\'{e}rsic profile, while the nucleated dwarfs are approximated by two components.~The robustness of the fitting procedure is estimated from a few systematically differing initial guesses for all model parameters.~For 88 non-nucleated dwarfs, all varying initial guesses immediately converged to the same solution and are thus considered to be robust fits.~On the other hand, for 114, mostly fainter objects, {\sc galfit} could not find a stable solution, so we refine the fitting by adopting the following iterative procedure.~First, we estimate the total galaxy luminosity from {\sc SExtractor} MAG\_AUTO values using the corresponding segmentation maps.~These magnitudes are then used as initial guesses for the {\sc galfit} S\'{e}rsic models and kept fixed during the fit.~In this way, the fits are stabilized so that the remaining model parameters can be derived.~Next, the newly estimated parameters are kept fixed and the galaxy luminosity is recomputed with {\sc  galfit}.~In the final step of the procedure, the newly determined magnitude is again fixed to recompute the other parameters.~In this way, all non-nucleated dwarfs were successfully fit, and each galaxy model and residual image were visually inspected to ensure the derived parameters are statistically robust and provide reliable approximations. 

To achieve consistent and meaningful fits for the nucleated dwarfs, we modified the procedure to properly account for the excess light in the nuclei.~We first attempted to fit a galaxy model with two components, e.g.\ a Nuker or King profile for the nucleus and a S\'{e}rsic profile for the extended, diffuse galaxy spheroid. However, very few stable solutions were found for the nuclei due to their small sizes. As a solution to this problem, we instead mask the nuclei and then fit a single S\'{e}rsic to the outer diffuse galaxy spheroid.~From the residual images (including the galaxy nuclei) we generate improved masks for the nuclei and any other sources contaminating the diffuse components of the dwarfs.~This procedure resulted in stable fits for the diffuse spheroids of all nucleated dwarf galaxy candidates.~Again, the residual and model images for each galaxy were visually inspected to validate the robustness of the computed model parameters.

For 20 galaxies in our sample we could not obtain stable fits mostly due to their locations near bright stars or simply being too faint.~All remaining 264 dwarfs can be reliably fit with a S\'{e}rsic model with $0\!\lesssim\!n\!\lesssim\!2$ and show a well-defined peak around $n\!\approx\!0.7$.~This is consistent with previous measurements \citep[e.g.][]{cot07} which show that the main bodies of galaxies are well defined by a S\'{e}rsic profile with S\'{e}rsic indices that vary smoothly from $n\!\approx\!4$ (deVaucouleurs $r^{1/4}$ profile) in the most massive systems to $n\!\approx\!1$ (exponential profiles) in the lowest mass galaxies.

\section{Discussion}
\label{sec:disc}

In this section we explore various parameter spaces of the newly expanded Fornax dwarf galaxy population.~We first compare them to other galactic systems in the context of their respective size-luminosity relations, and then provide an overall view of their spatial distributions with respect to both each other, and the population of giant elliptical galaxies in the Fornax cluster core.

\subsection{Size-Luminosity Relation}

Figure~\ref{dwarfsequence} shows the effective radii vs.\ absolute $i$-band magnitudes for our NGFS sample dwarf galaxies in relation to other stellar systems in the nearby Universe.~The detected Fornax dwarfs range mainly between values of $\langle\mu_i\rangle_{\rm{e}}\!\approx\!22\!-\!28$ mag arcsec$^{-2}$, with absolute luminosities down to approximately $-8$ mag -- similar to the Virgo ultra-faint dwarf (UFD) recently detected by \cite{jan14} -- and half-light radii from $r_e\!\simeq\!90$\,pc up to $\sim\!2.5$\,kpc.~For the 92 galaxies in common between our NGFS and the FCC catalogue, we find good agreement between the magnitudes and effective radii, confirming the robustness of our measurements.~The measured luminosities and structural parameters are provided in Table~\ref{tab:dwarftable}.

At the bright end of the NGFS sequence in Figure~\ref{dwarfsequence} we find dwarf galaxies that resemble ultra-diffuse galaxies identified by \cite{mih15} in Virgo and \cite{dokkum15a} in the Coma cluster.~We therefore confirm that UDGs also exist in the central regions of the Fornax galaxy cluster and are not confined to massive galaxy clusters alone.~However, at a given luminosity the Fornax dwarfs are significantly smaller than the Coma and Virgo UDGs.~The existence of such galaxies has been predicted and observed by \cite{dal97a}.~At the faint end of the NGFS dwarf sequence, the lowest surface brightness dwarfs find analogues in Local Group galaxies such as the Draco, Ursa Major, and Andromeda XXVII dwarf spheroidals \citep{mcc12}.~Overall the NGFS dwarf sequence appears very similar to the size-luminosity relation of Local Group dwarf spheroidal galaxies.~However, within the photometric completeness limits of our data we detect a small offset towards a higher average surface brightness of NGFS dwarfs compared with Local Group dwarfs.~This might be due to either {\it i)} sample statistics of the Fornax dwarf population, {\it ii)} the harsher ram-pressure stripping and tidal environment in Fornax which at a given total luminosity can render galaxies more compact \citep{may06} or/and {\it iii)} different star formation histories and stellar population properties of NGFS dwarfs \citep[e.g.][]{gue15}.~We will discuss these relations in detail in a subsequent paper.

\subsection{Nucleation Fractions}

We divide our sample into nucleated and non-nucleated dwarf galaxies, which was an obvious classification for almost all galaxies.~Those very few galaxies, with initially uncertain classification were also easily assigned to either of the categories after subtraction of their spheroid light component.~We find that the nucleated dwarfs are on average significantly brighter than their non-nucleated counterparts.~Both nucleated and non-nucleated sub-samples clearly show a size-luminosity relation, in the sense of increasing size with increasing galaxy luminosity.~However, the slope of this trend is significantly shallower for the nucleated dwarfs.

We investigate how the nucleation fraction ($f_{\rm nuc}$) changes as a function of galaxy luminosity.~Figure~\ref{dwarfsequence} shows that $f_{\rm nuc}$ is a strong function of galaxy luminosity in the sense that fainter dwarfs show systematically lower $f_{\rm nuc}$.~We find that many FCC dwarfs had wrong nucleation flags and we revised their classification.~We note that FCC galaxies are, on average, brighter than our newly detected dwarfs, and that 45 out of 92 reclassified FCC galaxies show a nucleus, i.e.~$\sim\!49$\%, which is consequence of the higher spatial resolution\footnote{Note that \cite{cot06} find also a much higher nucleation fraction in Virgo compared to the Virgo Cluster Catalog \citep{bin85}.}.~For the newly detected NGFS dwarfs only 15 out of 158 ($\sim\!9$\%) are nucleated.~In any case, Figure~\ref{dwarfsequence} clearly shows a trend of increasing nucleation fraction with galaxy spheroid luminosity.~For luminosities fainter than $M_i\!\simeq\!-11$\,mag, we observe $f_{\rm nuc}\!\simeq\!3\%$, whereas for those galaxies brighter than $M_i\!\simeq\!-15$\,mag, $f_{\rm nuc}$ increases to $>\!75$\%.

From HST/ACS imaging analysis, \cite{tur12} find an average nucleus-to-galaxy luminosity ratio, $\langle\eta\rangle\!=\!{\cal L}_n/{\cal L}_g$, for bright Fornax early-type galaxies of 0.41\%,~which corresponds to a magnitude difference of 6 mag.~Considering our approximate point-source detection limit at $M_i\!\approx\!-8$ \,mag, we should not be detecting nuclei in galaxies fainter than $M_i\!\approx\!-14$\,mag, which we clearly do.~This indicates that for those intermediate luminosity galaxies, {\it i)} the galaxy spheroids are too faint for their nuclei, e.g.~due to tidal stripping of their halos, or {\it ii)} the nuclei are brighter than what would be expected from $\langle\eta\rangle$ of brighter early-type galaxies, i.e.~hosting younger and/or lower-metallicity stellar populations, or {\it iii)} the formation of the nuclei was more efficient, i.e~at higher redshifts when the environmental density was higher.~We will discuss this interesting result in an upcoming paper.

\subsection{Spatial Distribution}

A comprehensive view of the new dwarf candidate spatial distribution is illustrated in Figure~\ref{surfdense}, where the dwarfs have been limited to those within 60\arcmin\ ($\sim\!350$\,kpc) of NGC\,1399 to ensure uniform angular representation.~The result of a two-point angular correlation function \citep{landy93} analysis of the NGFS dwarf sample is shown in the top-left panel along with 1-$\sigma$ bounds.~On length scales below $\sim\!100$\,kpc, an intriguing excess appears, indicating the clustering of dwarf galaxies.~This may point to {\it i)} dwarf-giant galaxy clustering or/and {\it ii)} bound dwarf galaxy groups that survive in the galaxy cluster environment on those or smaller spatial scales.~Interestingly, such dwarf galaxy clustering has been reported in high-resolution numerical simulations \citep{kra04, wei08, vdaa14, whe15}.~We will investigate this result and its interpretation in a dedicated future paper.

The bottom-left panel of Figure~\ref{surfdense} shows the projected radial surface number density ($\Sigma_N$) profiles for the non-nucleated dwarfs (red), nucleated fraction (blue), and total population (brown).~The $\Sigma_N$ values show the number of dwarfs per arcmin$^2$ within adaptively sized annuli containing 20 galaxies each for the total sample, and appropriately scaled for the two subsamples such that similar total data points are obtained, i.e.~15 and 5 galaxies per bin for the non-nucleated and nucleated samples, respectively.~Cluster centric distances corresponding to annuli mid-points are shown along the top- and bottom axes in physical and angular units.~The non-nucleated sample and total population are both consistent with having flat $\Sigma_N$ distributions of $\sim\!0.01\!-\!0.02\,{\rm arcmin}^{-2}$ out to $\sim\!350\,{\rm kpc}$, whereas the nucleated dwarf population (noting the much smaller sample) appears to decline outside of $\sim\!200\,{\rm kpc}$.~However, this result requires confirmation with a larger sample from the full NGFS survey footprint.

The distributions in azimuthal angles are shown in the right-hand panel of Figure~\ref{surfdense}, where the color opacity has been decreased to show where the samples overlap.~The smooth curves indicate non-parametric Epanechnikov-kernel probability density estimates based on the discrete data, and each curve exhibits several distinct peaks.~For all curves, there is a noteworthy over density along the $\Phi\simeq100^\circ$ direction, which coincides with the inter-cluster medium between the giant galaxies NGC\,1427 and NGC\,1399 (see Figure~\ref{fig1}). The non-nucleated population appears much more stochastic, but with a notable broad distribution in the range $200^\circ\la\Phi\la300^\circ$, with individual peaks that closely follow the distributions of the other giant galaxies in the Fornax core, e.g.~NGC\,1386 ($\Phi\simeq220^\circ$), NGC\,1387, and 1379 ($\Phi\simeq250^\circ$), and NGC\,1374 ($\Phi\simeq275^\circ$). While the total number of nucleated dwarf candidates prohibit making definitive statements, their distribution tends to follow the same general East-West bimodality, but without the discrete over densities shown by the populous samples. Thus, while they are likely to trace the interactions between the dominant galaxies of Fornax, their broader distributions limit what can be concluded regarding specific associations at this point.~The full NGFS survey dataset will expand this result to a larger-scale analysis.

\acknowledgments

This project is supported by FONDECYT Postdoctoral Fellowship Project No.~3130750, FONDECYT Regular Project No.~1121005 and BASAL Center for Astrophysics and Associated Technologies (PFB-06).~P.E.~acknowledges support from FONDECYT Postdoctoral Fellowship Project No.~3130485.~K.A.M. acknowledges support from FONDECYT Postdoctoral Fellowship Project No.~3150599.~M.A.T.~acknowledges the financial support through an excellence grant from the ``Vicerrector\'ia de Investigaci\'on" and the Institute of Astrophysics Graduate School Fund at Pontificia Universidad Cat\'olica de Chile and the European Southern Observatory Graduate Student Fellowship program.~G.G.~acknowledges support from FONDECYT Regular Project No.~1120195.

This project used data obtained with the Dark Energy Camera (DECam), which was constructed by the Dark Energy Survey (DES) collaboration.

This research has made use of the VizieR catalogue access tool and the Aladin plot tool at CDS, Strasbourg, France.~We are grateful to Roberto Gonz\'alez, Francisco Valdes, and David James for helpful discussions.

{\it Facilities:} \facility{CTIO (4m Blanco/DECam)}.

\clearpage

\end{document}